\documentclass[11pt]{article}
\usepackage{amsmath}
\usepackage{amssymb}
\usepackage{amsthm}
\usepackage{fullpage}
\usepackage{newpxtext,newpxmath}
\usepackage{hyperref}
\usepackage{todonotes}
\usepackage{enumitem}

\usepackage{authblk}

\newcommand{\rc}{\operatorname{rc}}

\newcommand{\R}{\mathbb{R}}
\newcommand{\Q}{\mathbb{Q}}
\newcommand{\Z}{\mathbb{Z}}
\newcommand{\zero}{\mathbf{0}}
\newcommand{\unit}{\mathbf{e}}

\newtheorem{theorem}{Theorem}[section]
\newtheorem{lemma}[theorem]{Lemma}

\title{The relaxation complexity of the standard simplex is logarithmic}
\author[1]{Gennadiy Averkov}
\author[2]{Simon Keil}
\author[2]{Stefan Weltge}
\affil[1]{Brandenburg University of Technology Cottbus–Senftenberg}
\affil[2]{Technical University of Munich}
\date{}

\begin{document}

\maketitle

\begin{abstract}
    For a set $X$ of integer points, the relaxation complexity $\rc(X)$ is the smallest number of facets of any polyhedron $P$ such that $P \cap \Z^d = X$. In this paper, we focus on the case where $X$ is the discrete standard simplex $\Delta_d = \{\zero, \unit_1, \dots, \unit_d\}$. We show that $\rc(\Delta_d) = O(\log d)$ by an explicit, elementary construction.
    This improves upon the previously best-known upper bound $\rc(\Delta_d) = O(d / \sqrt{\log d})$ due to Aprile, Averkov, Di Summa, and Hojny (2024) and matches an asymptotic lower bound by Averkov and Schymura (2022).
\end{abstract}

\section{Introduction}
Given a set $X \subseteq \Z^d$ of integer points, we call a polyhedron $P \subseteq \R^d$ a \emph{relaxation} of $X$ if $P \cap \Z^d = X$. For a set $X$ of integer points, the \emph{relaxation complexity} $\rc(X)$ is the smallest number of facets of any relaxation of $X$. The relaxation complexity can be seen as the minimal number of inequalities needed in an integer programming formulation without auxiliary variables whose feasible (integer) points are precisely the points of $X$. This concept was introduced by Kaibel and Weltge~\cite{Kaibel2015, weltge2015sizes}, who studied relaxation complexities of integer points corresponding to well-known combinatorial optimization problems such as the traveling salesman problem.
For several of these problems, they established exponential lower bounds, showing that auxiliary variables are often necessary to obtain compact integer programming formulations.

More general notions of relaxation complexity have also been studied.
Given an additional set $Y \subseteq \Z^d$, a natural generalization is to consider the smallest number $\rc(X,Y)$ of facets of a polyhedron $P$ such that $X \subseteq P$ and $(Y \setminus X) \cap P = \emptyset$.
Clearly, we have $\rc(X) = \rc(X, \Z^d) \ge \rc(X,Y)$.
For subsets $X \subseteq \{0,1\}^d$, Jeroslow~\cite{Jeroslow1975} already studied $\rc(X,\{0,1\}^d)$ under the name \emph{index}.
Averkov, Hojny, and Schymura~\cite{Averkov2023a, Averkov2023b} presented approaches to compute $\rc(X,Y)$ for finite sets $X$ and $Y$, and discussed in which situations $\rc(X)$ can be approximated by $\rc(X,Y)$ for particular finite sets $Y$.
Moreover, they pointed out that the notion of $\rc(X,Y)$ generalizes concepts that are examined in other fields, such as machine learning (see, e.g., \cite{astorino2002polyhedral,baum1990learning,dundar2008polyhedral,manwani2010learning,orsenigo2007accurately}), social choice (see, e.g.,~\cite{kober2021improved,kurz2016dimension,taylor2008mathematics}), and cryptography (see, e.g.,~\cite{sun2014automatic, udovenko2021milp}).

While the relaxation complexity provides a basic measure of how compactly a set $X$ can be described by linear inequalities, its precise value is known only for very few choices of $X$, among them discrete hypercubes~\cite{Kaibel2015}, cross-polytopes and axis-aligned boxes~\cite{Averkov2023a}.
However, its value is unknown for the arguably simplest (full-dimensional) set of integer points, the discrete standard simplex $\Delta_d = \{\zero, \unit_1, \dots, \unit_d\}$, where $\zero$ denotes the all-zero vector and $\unit_i$ denotes the vector that has a $1$ in the $i$-th coordinate and $0$ elsewhere.
Clearly, since $\Delta_d = \{x \in \Z^d : x_1,\dots,x_d \ge 0, \sum_{i=1}^d x_i \le 1\}$, we have $\rc(\Delta_d) \le d + 1$.
In~\cite{Kaibel2015}, it was asked whether this bound is tight.

The work of Averkov and Schymura~\cite{Averkov2022} implies that $\rc(\Delta_d) = d + 1$ indeed holds for $d \leq 4$.
However, they were only able to prove a lower bound of $\rc(\Delta_d) = \Omega(\log d)$ for general $d$.
In fact, they showed that $\rc(X) = \Omega(\log d)$ holds for any finite $d$-dimensional set of integer points $X$.

For $d \ge 5$, suppose that $P$ is a relaxation of $\Delta_d$ with fewer than $d + 1$ facets.
This implies that $P$ is unbounded and hence must be defined by linear inequalities with irrational coefficients (otherwise $P$ would contain a rational ray and hence infinitely many integer points).
Somewhat surprisingly, Aprile, Averkov, Di Summa, and Hojny~\cite{Aprile2024} constructed such relaxations for $d \ge 5$, showing $\rc(\Delta_d) \le d$ and hence providing a natural example of a set of integer points for which the relaxation complexity is strictly smaller than its rational counterpart.
Moreover, they established the asymptotic bound $\rc(\Delta_d) = O(d / \sqrt{\log(d)})$, still leaving a gap to the lower bound by Averkov and Schymura.

In this paper, we close this gap by providing an explicit, elementary construction of a loga\-rithmic-size relaxation of $\Delta_d$.
Hence, we obtain the following result.
\begin{theorem}
    \label{thmMain}
    We have $\rc(\Delta_d) = \Theta(\log d)$.
\end{theorem}
Combining this with the general lower bound for finite $d$-dimensional sets by Averkov and Schymura, this also shows that the minimum of the relaxation complexity over all finite $d$-dimensional sets of integer points is $\Theta(\log d)$.

\section{Construction}

Our construction is easily formulated using the free join operation, which is frequently employed in the context of polyhedra (see, e.g., \cite[\S 15.1.3]{henk2017basic}) and allows us to decompose simplices into lower-dimensional ones.
For $X \subseteq \Z^n$ and $Y \subseteq \Z^m$, the \emph{free join} of $X$ and $Y$ is defined as
\[
    X \ast Y = \{(x, \zero, 0): x \in X\} \cup \{(\zero, y, 1): y \in Y\} \subseteq \Z^{n+m+1}. 
\]
We say that $A,B \subseteq \Z^d$ are \emph{unimodularly equivalent} if there exists a linear map $f : \R^d \to \R^d$ with $f(\Z^d) = \Z^d$ and $f(A) = B$.
Note that if $A$ and $B$ are unimodularly equivalent, then $\rc(A) = \rc(B)$.

\begin{lemma}
    \label{lemFreeJoinSimplex}
    The free join of $\Delta_n$ and $\Delta_m$ is unimodularly equivalent to $\Delta_{n + m + 1}$.
\end{lemma}
\begin{proof}
    Note that
    \[
        \Delta_n \ast \Delta_m = \{\zero, \unit_1, \dots, \unit_n\} \cup \{\unit_{n+m+1}, \unit_{n+1} + \unit_{n+m+1}, \dots, \unit_{n+m} + \unit_{n+m+1}\} \subseteq{\Z^{n+m+1}}.
    \]
    Let $f : \R^{n+m+1} \to \R^{n+m+1}$ be the linear map that maps $\unit_i$ to $\unit_i$ for $i \in \{1,\dots,n\} \cup \{n+m+1\}$ and $\unit_j$ to $\unit_j + \unit_{n+m+1}$ for $j \in \{n+1, \dots, n+m\}$. Then, $f(\Delta_{n+m+1}) = \Delta_n \ast \Delta_m$. Moreover, it is easy to see that $f$ is unimodular.
\end{proof}

Our construction is based on the following observation.

\begin{lemma}
    \label{lemFreeJoinRelaxation}
    If $X \subseteq \Z^d$ is finite, then $\rc(X \ast X ) \le \rc(X) + 4$.
\end{lemma}
\begin{proof}
    Let $P$ be a relaxation of $X$ and let $\alpha_1, \dots, \alpha_d \in \R$ be linearly independent over $\Q$.
    Define $g : \R^d \to \R$ by $g(x) = \sum_{i=1}^d \alpha_i x_i$ and set $M = \max_{x \in X} |g(x)|$.
    We claim that the polyhedron $Q$ given by
    \[
        Q = \{(x,y,h) \in \R^d \times \R^d \times \R : |g(x)| \le M (1-h), |g(y)| \le M h, x + y \in P\}
    \]
    is a relaxation of $X \ast X$.
    Since $Q$ has at most 4 more facets than $P$, this implies the statement of the lemma.

    Let $(x,y,h) \in X \ast X$. If $h = 0$, then $y = \zero$ and $x \in X$.
    By the choice of $M$, this implies $(x,y,h) \in Q$. 
    If $h = 1$, then $x = \zero$ and $y \in X$, and we again obtain $(x,y,h) \in Q$.
    Thus, $X \ast X \subseteq Q$.

    To see that $Q \cap \Z^{2d+1} \subseteq X \ast X$ holds, let $(x,y,h) \in Q \cap \Z^{2d+1}$.
    Since $|g(x)|, |g(y)|$ and $M$ are non-negative, the inequalities imply $0 \le 1 - h$ and $0 \le h$, which together with the integrality gives $h \in \{0,1\}$. If $h = 0$, then $|g(y)| \le M h$ implies $g(y) = 0$.
    Since the $\alpha_i$ are linearly independent over $\Q$ and $y$ is integer, this implies $y = \zero$.
    Thus, we have $x = x + y \in P$ and hence $(x,y,h) = (x, \zero, 0) \in X \ast X$.
    If $h = 1$, we analogously obtain $x = \zero$ and $y \in X$, which implies $(x,y,h) = (\zero, y, 1) \in X \ast X$.
\end{proof}

The proof of Theorem~\ref{thmMain} now follows from iteratively applying Lemmas~\ref{lemFreeJoinSimplex} and~\ref{lemFreeJoinRelaxation}, and the fact that $\rc(\Delta_d)$ is monotone in $d$ (see, e.g., \cite[Prop.~2]{Aprile2024}).

\section{Remarks}

\subsection{Asymptotic factor}

Let $c := \limsup_{d \to \infty} \rc(\Delta_d)/\log_2 d$ denote the asymptotic factor of the relaxation complexity of the standard simplex.
Averkov and Schymura~\cite{Averkov2022} showed that $c \ge 1$.
Our construction can be improved by considering the $k$-fold free join $X^{\ast k} = X \ast \dots \ast X$ of $X$ with itself.
Lemma~\ref{lemFreeJoinRelaxation} can be generalized by using the polyhedron
\[
    \{(x_1, \dots, x_k, h_1, \dots, h_k) \in \R^{kd} \times \R^k: |f(x_i)| \le M h_i \ \forall i \in [k], x_1 + \dots + x_k \in P, h_1 + \dots + h_k = 1 \}.
\]
In fact, assuming that $P$ is a relaxation of $X$, this
polyhedron yields a relaxation of $X^{\ast k}$ after eliminating the $h_1$ coordinate using the equality constraint $h_1 + \dots + h_k = 1$ and has at most $2k$ more facets than $P$.
Thus, we have $\rc(X^{\ast k}) \le \rc(X) + 2k$.
Using this with Lemma~\ref{lemFreeJoinSimplex}, we deduce that
\begin{equation}
    \label{eqRcMultiple}
    \rc(\Delta_{dk - 1}) = \rc(\Delta_{d-1}^{\ast k}) \le \rc(\Delta_{d-1}) + 2k.
\end{equation}
Iterating this $\ell$ times, we obtain $\rc(\Delta_{dk^\ell - 1}) \le \rc(\Delta_{d-1}) + 2 k \ell$. In particular, we have
\[
    \rc(\Delta_{k^\ell - 1}) \le \rc(\Delta_{k-1}) + 2 k (\ell-1) \le k + 2k (\ell-1) = (2\ell - 1)k \le 2\ell k.
\]
Setting $\ell = \lceil \log_k (d+1) \rceil$ and using the monotonicity of $\rc(\Delta_d)$, we conclude
\begin{equation}
    \label{eqRcGeneral}
    \rc(\Delta_d) \le 2k\cdot \lceil \log_k (d+1) \rceil.
\end{equation}
Choosing $k = 3$ hence yields $c \le 6 \ln(2)/\ln(3) \approx 3.79$.
Determination of the exact value of $c$remains an open question.

\subsection{Degree of field extension}

Aprile et al.~\cite{Aprile2024} proposed to study the relaxation complexity $\rc_F$ with respect to arbitrary subfields $F$ of $\R$, i.e., the minimal size of a relaxation that only uses coefficients from $F$. They remarked that their construction works for every field extension over $\Q$ of degree at least $d+1$.
The construction above shows that the bound $\rc(\Delta_d) \le 6 \cdot \lceil  \log_3 (d+1) \bigr \rceil$ obtained in \eqref{eqRcGeneral} holds for all $F$ with $\dim_\Q(F) \ge d/3$. If we consider a field of extension degree $\delta := \dim_\Q(F) <  d/3$ and additionally use \eqref{eqRcMultiple}, we obtain
\[
    \rc_F(\Delta_d) \le \rc_F(\Delta_{\lceil (d+1)/(3\delta)\rceil 3 \delta - 1}) \le \rc_F(\Delta_{3\delta-1}) + 2 \cdot \left\lceil \frac{d+1}{3\delta} \right\rceil \le 6 \cdot \lceil  \log_3 (3 \delta) \bigr \rceil + 2 \cdot \left\lceil \frac{d+1}{3\delta} \right\rceil.
\]
Note that for $\delta = \Omega(d/\log d)$, the right hand side is of order $O(\log d)$, meaning that the relaxation complexity is logarithmic even for a sublinear degree of the field extension. Moreover, for $\delta = \Omega(\sqrt{\log(d)})$, the right hand side is of order $O(d/\sqrt{\log(d)})$, which matches the bound obtained by Aprile et al.~\cite{Aprile2024}, but reduces the required dimension of the field extension.

\bibliographystyle{plain}
\bibliography{references}

\end{document}